\documentclass[preprint,12pt]{elsarticle}
\usepackage{amsthm}
\usepackage{amsmath}
\usepackage{amssymb}
\usepackage{amstext}
\usepackage{bm}
\usepackage{graphicx}
\usepackage{epstopdf}
\usepackage{subcaption}

 \newcommand{\bl}{\big<}
  \newcommand{\bg}{\big>}
  \newcommand{\eps}{\varepsilon}
 
\biboptions{sort&compress}

\journal{ArXiv.org}

\begin{document}

\begin{frontmatter}

\title{Nonclassical Particle Transport in the 1-D Diffusive Limit}

\author[ucb]{Richard Vasques\corref{cor1}}
\author[ucb]{Rachel Slaybaugh}
\author[aachen]{Kai Krycki}
\cortext[cor1]{Corresponding author: richard.vasques@fulbrightmail.org\\
Dept. of Nuclear Engineering, University of California, Berkeley, 4103 Etcheverry Hall, MC 1730, Berkeley, CA 94720-1730}
\address[ucb]{Dept. of Nuclear Engineering, University of California, Berkeley}
\address[aachen]{Aachen Institute for Nuclear Training GmbH, Aachen, Germany}







\end{frontmatter}

\section{Introduction}\label{sec1}
\setcounter{section}{1}
\setcounter{equation}{0} 

A \textit{nonclassical linear Boltzmann equation} has been recently proposed  to address nonexponential attenuation of the particle flux  in certain inhomogeneous random media applications \cite{vas14a}. In particular, this effect arises in Pebble Bed reactor cores, in which the locations of the pebbles are spatially correlated \cite{vas14b}. 

In this paper, we investigate nonclassical particle transport taking place in a 1-D random periodic diffusive system. We provide computational results that validate the theoretical predictions, demonstrating for the first time that the solution of the nonclassical particle transport equation is well-approximated by the solution of the nonclassical diffusion equation. 

For simplicity, we consider the following assumptions: (i) transport occurs in \textit{rod geometry}, in which particles can only move in the directions $\mu=\pm 1$; (ii) transport is monoenergetic; and (iii) scattering is isotropic. In this case, the 1-D nonclassical linear Boltzmann equation is written as
\begin{align}
\label{1}
&\frac{\partial\psi^\pm}{\partial s}(x,s) \pm\frac{\partial\psi^\pm}{\partial x} (x,s) + \Sigma_t(s)\psi^\pm(x,s)\\
&\,\,= \frac{\delta(s)}{2}\left[ c\int_0^\infty \Sigma_t(s')\left[\psi^\pm(x,s') + \psi^\mp(x,s')\right]ds'+ Q(x) \right], \nonumber
\end{align}
where $x =$ position, $s =$ the path-length traveled by the particle since its previous
interaction (birth or scattering), $\psi^\pm$ is the nonclassical angular flux in the directions $\pm 1$, $c$ is the scattering ratio (probability of scattering), and $Q(x)$ is an isotropic internal source. The function $\Sigma_t(s)$ represents the collision probability (ensemble-averaged over all possible physical realizations of the system), such that
\begin{equation}
\Sigma_t(s)ds = \begin{array}{l}
\text{the probability that a particle, scattered or}\\
\text{born at any point $x$, will experience a}\\
\text{collision between $x + s$ and $x + (s+ds)$.}
\end{array} \nonumber
 \end{equation}
In this situation, the probability density function for a particle's distance-to-collision is given by
\begin{equation}\label{2}
p(s) = \Sigma_t(s)e^{-\int_0^s \Sigma_t(s')ds'}, 
\end{equation}
such that its $m^{th}$ moment is defined as
\begin{align}
\bl s^m \bg = \int_0^{\infty}s^mp(s)ds\,.\nonumber
\end{align}
If $\Sigma_t(s) = \Sigma_t = $ constant (classical total cross section), we obtain the exponential
\begin{equation}
p(s) = \Sigma_t e^{-\Sigma_ts},\label{3}
\end{equation}
and Eq.\ \eqref{1} reduces to the classical linear Boltzmann equation
\begin{equation}
\pm \frac{\partial \Psi^\pm}{\partial x}(x) + \Sigma_t \Psi^\pm(x) = \frac{\Sigma_s}{2}\left[\Psi^\pm(x)+\Psi^\mp(x)\right]+ \frac{Q(x)}{2} \nonumber
\end{equation}
for the classical angular flux 
\begin{equation}
\Psi^\pm(x) = \int_0^\infty \psi^\pm(x,s)ds. \nonumber
\end{equation} 

\section{Asymptotic Analysis}\label{sec2}
\setcounter{section}{2}
\setcounter{equation}{0} 

Following \cite{vas14a},  we scale $\Sigma_t = O(1)$, 
$ 1-c = O(\varepsilon^2) $, $Q=O(\varepsilon^2)$,  $\partial \psi / \partial s = O(1)$, and $\partial \psi/\partial x = O(\eps)$, with $\varepsilon \ll 1$. In this scaling, Eq.\ \eqref{1} yields
  \begin{align}
    &\frac{\partial \psi^\pm}{\partial s}  (x,s) 
      \pm \eps\frac{\partial\psi^\pm}{\partial x}(x, s)
       + \Sigma_t(s) \psi^\pm( x, s)   \label{4}\\
   & \,\,= \delta(s)\frac{1-\eps^2(1-c)}{2}\int_0^{\infty}\Sigma_t(s') 
      \left[\psi^\pm(x, s')+\psi^\mp(x,s')\right]ds'  \nonumber \\
     &\,\,\,\,\,\,\,\, + \varepsilon^2 \delta(s)\frac{Q(x)}{2} \,.
 \nonumber
  \end{align}
Let us define $\hat\psi^\pm(x, s)$ such that
  \begin{align}
   \psi^\pm( x,s) &\equiv 
          \hat\psi^\pm(x,s) \frac{e^{-\int_0^s \Sigma_t(s') ds'}}{\bl s\bg}\,.\nonumber
  \end{align}
Then, using Eq.\ \eqref{2}, Eq.\ \eqref{4} becomes the following equation for $\hat\psi^\pm(x, s)$:
  \begin{align}
    &\frac{\partial \hat\psi^\pm}{\partial s} (x,s) 
      \pm \varepsilon \frac{\partial\hat\psi^\pm}{\partial x}(x, s) \nonumber \\
   & \,\,= \delta(s)\frac{1-\eps^2(1-c)}{2} \int_0^{\infty} \left[
      \hat\psi^\pm(x, s')+\hat\psi^\mp(x,s')\right] p(s') \, ds'\nonumber\\
      & \,\,\,\,\,\,\,\,
      + \varepsilon^2 \delta(s) \bl s\bg \frac{Q(x)}{2} \,.\nonumber 
  \end{align}
This equation is mathematically equivalent to:
   \begin{subequations}\label{5}
   \begin{align}
      &\frac{\partial \hat\psi^\pm}{\partial s} (x,s) 
         \pm \varepsilon \frac{\partial\hat\psi^\pm}{\partial x}(x, s) = 0 \, \quad s > 0 \,,\label{5a}
         \end{align}
         and
         \begin{align}
       &\hat\psi^\pm(x, 0)  =\\
       &\,\,\,\,=\frac{1-\eps^2(1-c)}{2} \int_0^{\infty} p(s')
\left[\hat\psi^\pm(x ,s')+\hat\psi^\mp(x ,s')\right] ds'  \nonumber\\
         &\,\,\,\,\,\,\,\,\,\, + \varepsilon^2 \bl s\bg \frac{Q(x)}{2} \,,\nonumber
   \end{align}
   \end{subequations}
where $\hat\psi^\pm(x,0) = \hat\psi^\pm(x,0^+)$. Integrating Eq.\ \eqref{5a} over $0 < s' < s$, we obtain:
   \begin{align}
      &\hat\psi^\pm( x, s)  = \hat\psi^\pm(x, 0) \pm \varepsilon \frac{\partial}{\partial x} \int_0^s \hat\psi^\pm(x, s') \, ds' \nonumber\\
      & \,\, = \frac{1-\eps^2(1-c)}{2} \int_0^{\infty} p(s')
\left[\hat\psi^\pm(x,s')+\hat\psi^\mp(x, s')\right] ds'  \nonumber\\
         &\,\,\,\,\,\,\, + \varepsilon^2 \bl s\bg \frac{Q(x)}{2} \mp \varepsilon\frac{\partial}{\partial x} \int_0^s \hat\psi^\pm(x, s') \, ds' \,.\nonumber
   \end{align}
Introducing into this equation the ansatz 
   \begin{equation}
 \hat\psi^\pm(x, s) =  \sum_{n=0}^{\infty} \varepsilon^n
       \hat\psi_n^\pm(x, s) \nonumber
       \end{equation}
and equating the coefficients of different powers of $\varepsilon$, we obtain for $n \ge 0$:
   \begin{align}
&      \hat\psi_n^\pm(x, s)  = \frac{1}{2}\int_0^{\infty} p(s')\left[
\hat\psi_n^\pm(x, s')+\hat\psi_n^\mp(x,s')\right] ds' \label{6}\\
&\,\,\, \mp \frac{\partial}{\partial x} \int_0^s \hat\psi_{n-1}^\pm(x, s') \, ds' \nonumber \\
& \,\,\,\,\,\,\,\,\, -\frac{1-c}{2}\int_0^{\infty} p(s')\left[
\hat\psi_{n-2}^\pm(x, s')+\hat\psi_{n-2}^\mp(x,s')\right] ds'  \nonumber\\   
      & \,\,\,\,\,\,\,\,\,\,\,\,\,\,+ \delta_{n,2} \bl s\bg \frac{Q( x)}{2} \,, 
   \nonumber
   \end{align}
with $\hat\psi_{-1}^\pm=\hat\psi_{-2}^\pm=0$.

 Equation \eqref{6} with $n=0$ has the general solution
   \begin{equation}
      \hat\psi_0^\pm(x, s) = \frac{\hat\phi_0(x)}{2} \,, \nonumber
   \end{equation}
where $\hat\phi_0(x)$ is undetermined at this point. For $n=1$,
Eq.\ \eqref{6} has a particular solution of the form:
    \begin{equation}
      \hat\psi_{part}^\pm(x,s) = \mp \frac{s}{2}\frac{d \hat\phi_0}{d x}(x) \,,\nonumber
   \end{equation}   
and its general solution is given by 
   \begin{equation}
      \hat\psi_1^\pm( x, s) =  \frac{1}{2}\left[\hat\phi_1^\pm( x) \mp s\frac{d \hat\phi_0}{d x}(x)\right] \,,\nonumber
   \label{eq17}
  \end{equation}  
where $\hat\phi_1(x)$ is undetermined.

 Equation \eqref{6} with $n=2$ has a solvability condition, which is obtained by adding the equations for $\psi_2^+$ and $\psi_2^-$ and operating on them by $\int_0^{\infty} p(s) ( \cdot ) ds $; the solvability condition yields
   \begin{align}
      0 &= \frac{\bl s^2\bg}{2}\frac{d^2\hat\phi_0}{dx^2}(x) - (1-c)\hat\phi_0( x) + \bl s\bg Q( x)\,.\nonumber
   \end{align}
We can rewrite this equations as
\begin{align}
      -\frac{\bl s^2\bg}{2\bl s\bg}\frac{d^2\hat\phi_0}{dx^2}(x) + \frac{1-c}{\bl s\bg } \hat\phi_0(x) = Q(x)\,,\label{7}
      \end{align}
which is the nonclassical diffusion equation for Eq.\ \eqref{1}.

Therefore, the solution $\psi^\pm(x, s)$ of Eq.\ \eqref{4} satisfies
   \begin{equation}
      \psi^\pm(x, s) = \frac{\hat\phi_0(x)}{2} \frac{e^{- \int_0^s \Sigma_t( s') ds'}} {\bl s\bg} + O(\varepsilon) \,,\label{8}
   \end{equation} 
where $\hat\phi_0(x)$ satisfies Eq.\ \eqref{7}. The classical angular flux can be obtained to leading order by integrating Eq.\ \eqref{8} over $0 < s < \infty$:
   \begin{align}
   \Psi^\pm(x) = \int_0^{\infty}\psi^\pm(x,s)ds = \frac{\hat\phi_0(x)}{2}+ O(\varepsilon) \,.
   \nonumber
   \end{align}

As expected, if $p(s)$ is given by Eq.\ \eqref{3}, $\bl s\bg = 1/\Sigma_t$, $\bl s^2\bg = 2/\Sigma_t^2$, and it is easy to verify that Eq.\ \eqref{7} reduces to the classical diffusion equation
\begin{align}
 -\frac{1}{\Sigma_t}\frac{d^2\hat\phi_0}{dx^2}(x) + \Sigma_a \hat\phi_0(x) = Q(x)\,.\nonumber
\end{align}

\section{The 1-D Random Periodic System}\label{sec3}
\setcounter{section}{3}
\setcounter{equation}{0} 

Let us consider an \textit{infinite rod} consisting of periodically arranged solid and void layers of equal width, given by $\ell=1$.
We are interested in a \textit{finite random periodic system} with total width given by $2X = 2\ell M$, where the integer $M$ (the total length of each material in the system) satisfies $M = \varepsilon^{-1}$. Random realizations of this system can be obtained by randomly selecting any continuous $2X$ segment of the infinite rod described above. Vacuum boundary conditions are assigned at $x=\pm X$. 

In this finite random periodic system, the cross sections and source are stochastic functions of space. For the numerical results provided in this paper, we define the parameters at each spatial point $x$ in the system by
\begin{align}
\Sigma_t(x) &= \left\{\begin{array}{cc} 1, & \text{if $x$ is in solid} \\ 0, & \text{if $x$ is in void} \end{array}\right. ,\nonumber\\
Q(x) &= \left\{\begin{array}{cc} 2M^{-2}, & \text{if $x$ is in solid} \\ 0, & \text{if $x$ is in void} \end{array}\right. \nonumber,
\end{align}
and the absorption ratio by
\begin{align}
1-c = 1M^{-2}.\nonumber
\end{align}
These parameters are in agreement with the assumptions of our asymptotic analysis: $\Sigma_{t}$ is $O(1)$, $(1-c)$ and $Q$ are $O(\varepsilon^2)$, and the system is optically thick with $2X$ being $O(1/\varepsilon)$.
As $M$ increases, $\varepsilon$ decreases, and the 1-D system approaches the diffusive limit.

\section{Numerical Results}\label{sec4}
\setcounter{section}{4}
\setcounter{equation}{0} 

\begin{figure} 
  \centering
  \includegraphics[scale=1]{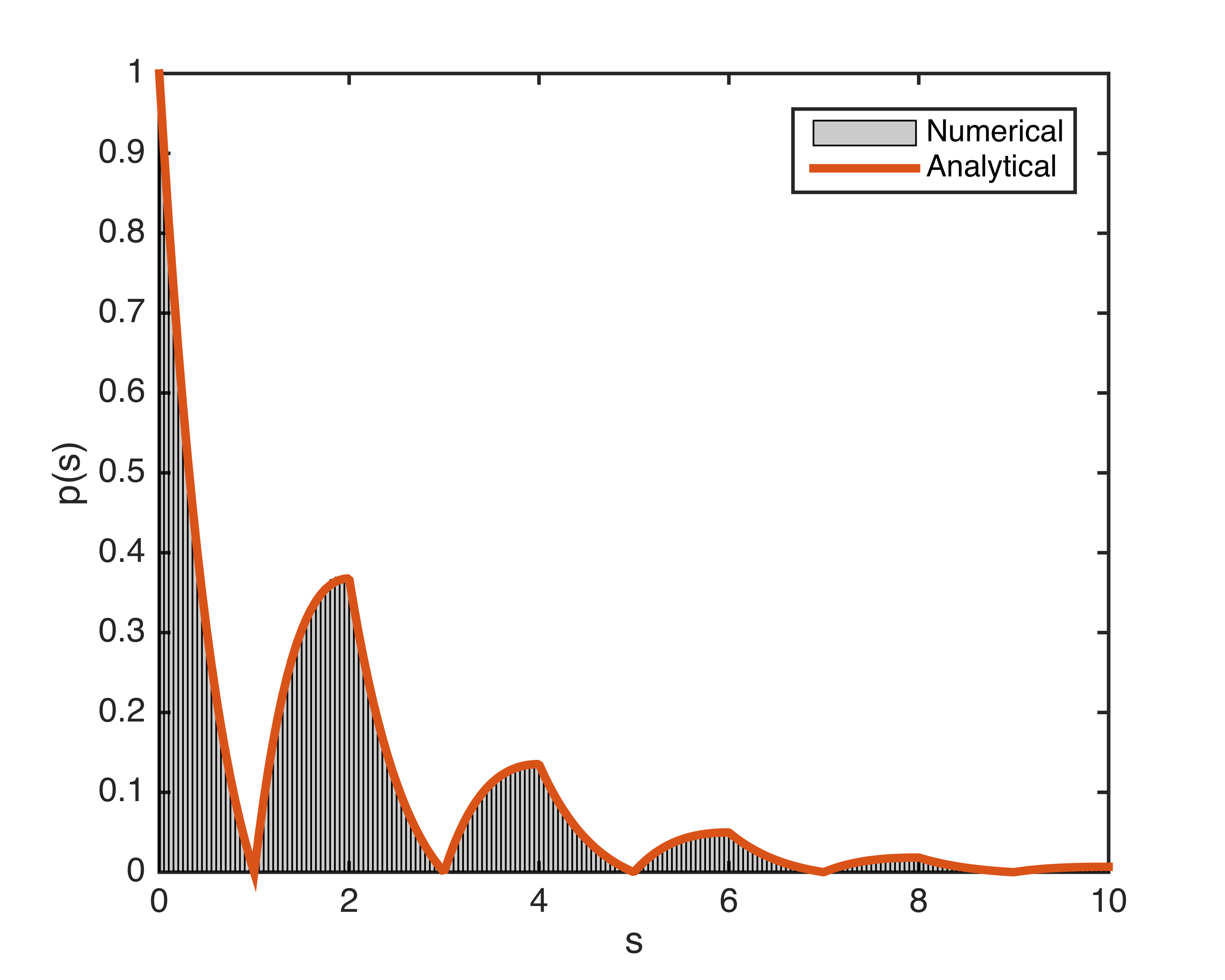}
  \caption{Probability density function for distance-to-collision.}
  \label{fig1}
\end{figure}

An analytical expression for $p(s)$ can be obtained for this type of 1-D system \cite{vas15}. Taking into account the parameters considered in this paper, the path length distribution function is given by 
\begin{align}
p(s) = \left\{
\begin{array}{ll}
(2n+1-s)e^{-(s-n)}, & \text{\hspace{-3pt}if $2n\leq s \leq 2n+1$}\\
(s-2n-1)e^{-(s-n-1)}, & \text{\hspace{-3pt}if $2n+1\leq s \leq 2(n+1)$}
\end{array}
\right.\nonumber
\end{align}
for $n=0,1,2,...$ (see Figure \ref{fig1}); this yields the moments
\begin{align}
\bl s\bg &= 2,\nonumber\\
\bl s^2\bg &= 5+\frac{2e}{e-1} \approx 8.1640 \, .\nonumber
\end{align}
The ensemble-averaged collision probability can be written as \cite{vas14a}
\begin{align}
\Sigma_t(s)ds = \frac{p(s)ds}{1-\int_0^sp(s')ds'}\,.\nonumber
\end{align} 

Finally, since the volume fraction of solid and void materials are the same in any realization of the system, the ensemble-averaged source is simply given by
\begin{align}
\bl Q\bg = \frac{Q(x)}{2} = 1M^{-2}.\nonumber
\end{align}


Following the procedure described in \cite{vas15}, we rewrite Eq.\ \eqref{1} in its initial value form as
\begin{subequations}\label{9}
\begin{align}
&\frac{\partial\psi^{\pm}}{\partial s}(x,s) \pm \frac{\partial \psi^{\pm}}{\partial x}(x,s) + \Sigma_t(s)\psi^{\pm}(x,s)  = 0, \label{9a}\\
& \psi^{\pm}(x,0)= \frac{c}{2} \int_0^\infty \Sigma_t(s')[\psi^{+}(x,s')+\psi^{-}(x,s')]ds' \label{9b}\\
& \hspace{50pt}+ \frac{\bl Q\bg}{2}. \nonumber
\end{align}
\end{subequations}

We then adapt the HLL finite volume method introduced in \cite{kry13}; this method is of first order in the pseudo-time variable $s$ and in the spatial variable $x$. In our calculations, we have cut off the integration at $s_{\text{max}}=45$, and have chosen the trapezoidal rule. We have used the mesh interval $\triangle x=1/400$ (= smallest  $\eps^2$ considered in this paper), and a CFL number $0.5$ (that is, $\triangle s = 1/800$). 

This system is solved in a source-iteration manner, where we iterate between the discretized forms of Eq.\ \eqref{9b} and Eq.\ \eqref{9a}. The ensemble-averaged nonclassical scalar flux is given by $\Phi(x) = \int_0^{45}[\psi^+(x,s)+\psi^-(x,s)]ds$.

We consider three different problems, with $M=10, 15,$ and $20$. As $M$ increases, we expect the solution of Eq.\ \eqref{1} to converge to the
solution of the nonclassical diffusion formulation given by Eq.\ \eqref{7}.
It was shown in \cite{kry13} that the contraction
rate for the source iteration is given by the scattering ratio $c$. To converge the solution, 299 source
iterations were needed for $c= 0.99$ ($M=10$); 579 iterations for $c = 224/225\approx 0.9956$ ($M=15$); and 936 iterations for $c= 0.9975$ ($M=20$).

Since our asymptotic analysis does not include boundary conditions, we solve Eq.\ \eqref{7} using the extrapolated endpoint boundary conditions, taking the extrapolation distance to be the diffusion coefficient:
\begin{align}
\hat\phi_0\left(X+\frac{\bl s^2\bg}{2\bl s \bg}\right) = \hat\phi_0\left(-X-\frac{\bl s^2\bg}{2\bl s\bg}\right)=0\,.\nonumber
\end{align}

\begin{figure} 
  \centering
  \includegraphics{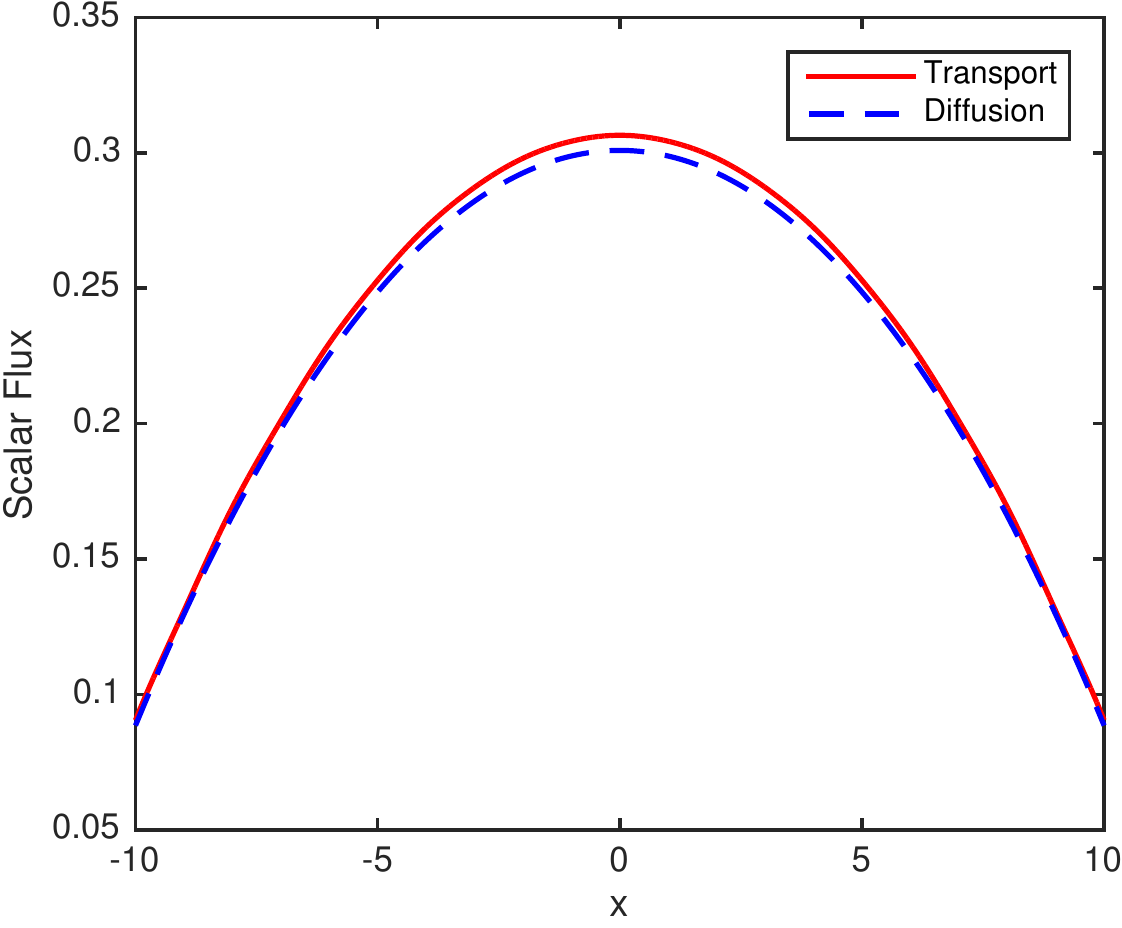}
  \caption{Scalar fluxes obtained by solving the nonclassical transport and the nonclassical difusion equations with $M=10$.}
  \label{fig2}
\end{figure}

\begin{figure} 
  \centering
  \includegraphics{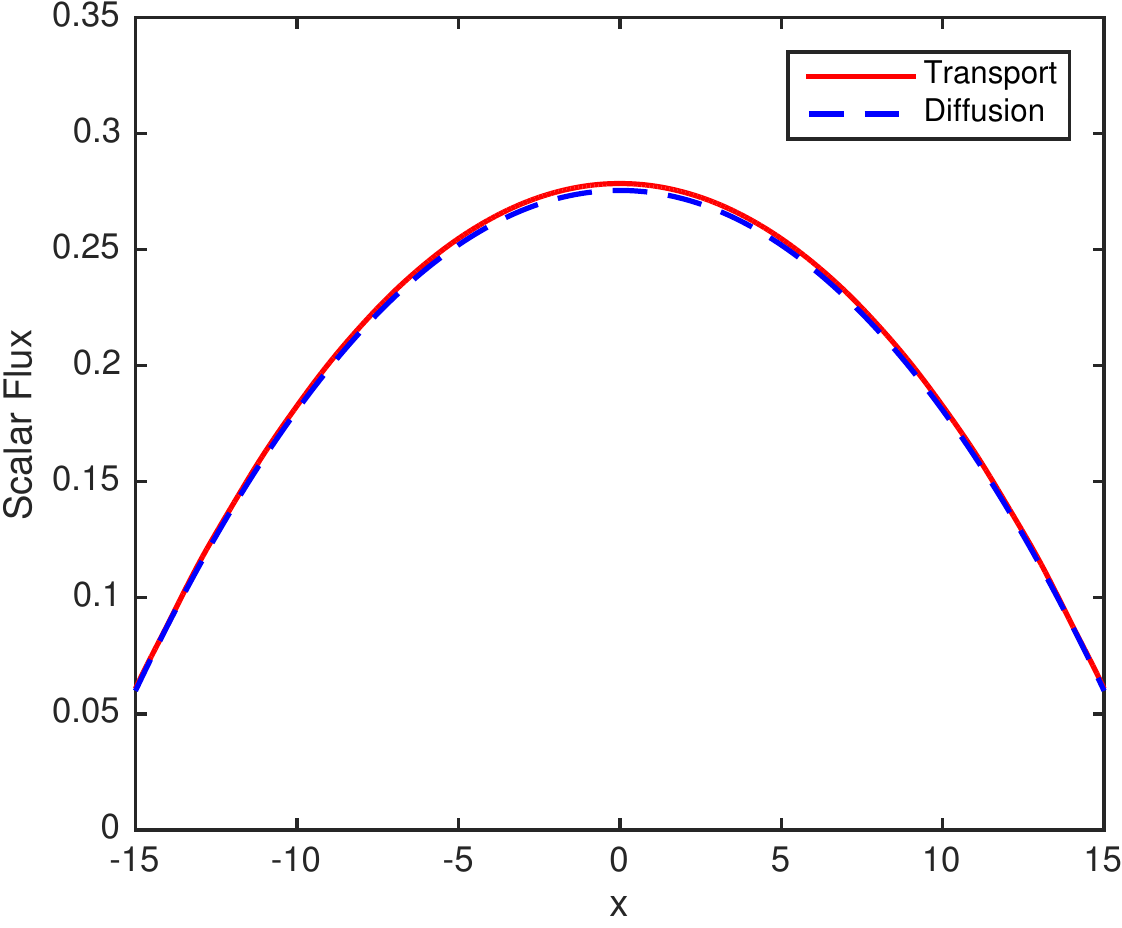}
  \caption{Scalar fluxes obtained by solving the nonclassical transport and the nonclassical difusion equations with $M=15$.}
  \label{fig3}
\end{figure}

\begin{figure} 
  \centering
  \includegraphics{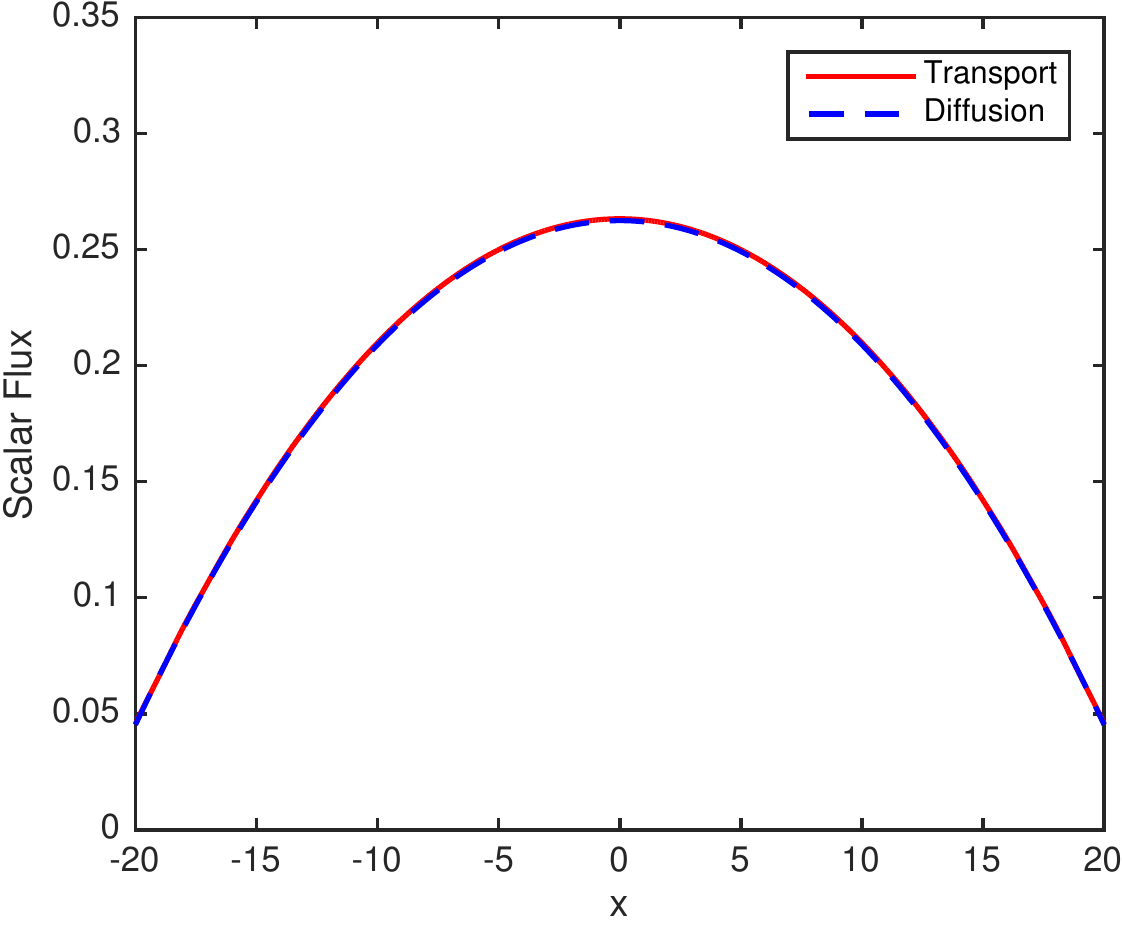}
  \caption{Scalar fluxes obtained by solving the nonclassical transport and the nonclassical difusion equations with $M=20$.}
  \label{fig4}
\end{figure}

As anticipated, the solutions plotted in Figures \ref{fig2}, \ref{fig3}, and \ref{fig4} confirm our claim: the nonclassical scalar flux and the nonclassical \textit{diffusion} scalar flux increasingly agree as $M$ increases and the system approaches the diffusive limit.

\section{Conclusions}\label{sec5}
\setcounter{section}{5}
\setcounter{equation}{0} 

This paper provides numerical results that demonstrate the validity of the nonclassical diffusion approximation to the nonclassical transport equation in certain 1-D diffusive systems. To our knowledge, this is the first time computational results that validate the predictions of the asymptotic theory for the nonclassical model are presented. This result provides a more solid foundation 
in which to improve this theory for relevant nuclear applications. 

\section*{Acknowledgments}
This paper was prepared by Richard Vasques and Rachel Slaybaugh under award number NRC-HQ-84-14-G-0052 from the Nuclear Regulatory Commission. The statements, findings, conclusions, and recommendations are those of the authors and do not necessarily reflect the view of the U.S.\ Nuclear Regulatory Commission.


\pagebreak



\end{document}